**Spin and Magnetism in 2D Materials**
Roland K. Kawakami, The Ohio State University

*Status.* Two-dimensional (2D) materials provide a unique platform for spintronics and magnetism, where the atomic thinness of the layers leads to strong tunability via electrostatic gates, as discussed by S. O. Valenzuela in the 2017 Magnetism Roadmap. Various types of 2D materials contribute distinct spin-dependent properties (Fig. 1): graphene provides excellent spin transport [1], transition metal dichalcogenides (TMDCs: $MX_2$, with M = Mo, W and X = S, Se) provide strong spin-orbit coupling and valley-selective optical transitions [2] and 2D magnets provide non-volatile storage and capabilities for spin filtering, injection, and detection [3]. By combining these materials in stacked van der Waals (vdW) heterostructures, their various properties are integrated within a single structure. Beyond the simple addition of functionalities, quantum mechanical interactions across interfaces produce spin proximity effects where properties of 2D layers are altered by imprinting characteristics of neighboring layers. These properties enable potential applications in efficient non-volatile memory, spin-based logic, and spin-dependent optoelectronics.

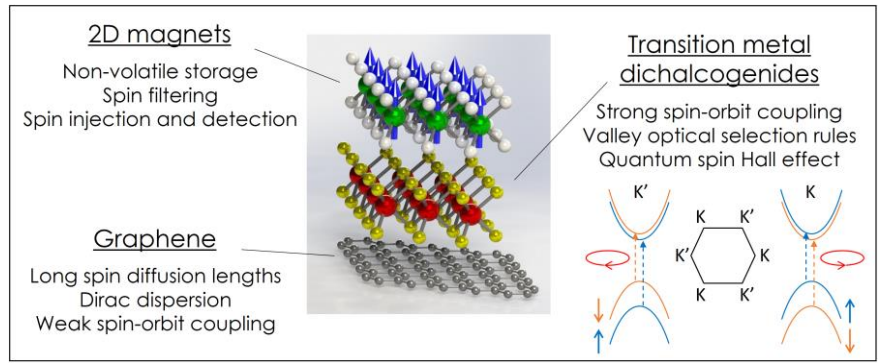

**Figure 1.** 2D materials for spintronic heterostructures.

Graphene exhibits the longest room temperature spin diffusion length (~30 µm) of any material, but the weak spin-orbit coupling has limited its capabilities for spin-charge conversion and electrical manipulation of spin [1]. Stacking a TMDC layer onto graphene imparts a proximity spin-orbit coupling, which was most convincingly demonstrated through spin precession experiments on $MoSe_2$/graphene and $WS_2$/graphene spin valves [1]. Fig. 2a [4] shows oblique spin precession measurements on a $WS_2$/graphene spin valve, where the dependence of the spin signal on the B-field angle displays a highly non-linear dependence (green data). This indicates a much longer spin lifetime for out-of-plane spins vs. in-plane spins, which is a smoking-gun indicator of proximity spin-orbit coupling in graphene induced by the $WS_2$. Subsequently, proximity spin-orbit coupling in TMDC/graphene heterostructures were used to demonstrate spin-charge conversion by spin Hall and Rashba-Edelstein effects, using spin precession to avoid spurious signals [1]. Meanwhile, electrical control of spin transport and spin relaxation was demonstrated. Control of spin transport by electric gates was achieved using a graphene spin valve with $MoS_2$ on top [1]. Fig. 2b [5] shows the increase of conductivity (black curve) of the n-type $MoS_2$ with gate voltage ($V_g$). This increases the spin absorption from graphene to $MoS_2$, which shunts away spin current from the graphene, eventually leading to zero spin current ($\Delta R_{NL} = 0$) for $V_g > 15$ V. In addition, electric control of spin relaxation was achieved in gated bilayer graphene, surprisingly without the need for proximity spin-orbit coupling [1]. Applying a perpendicular electric field opens up a band gap and the intrinsic spin-orbit splitting, though small (~24 µeV), produces an out-of-plane spin-orbit field to strongly increase the out-of-plane spin lifetime while decreasing the in-plane spin lifetime. This was identified through oblique spin precession measurements on bilayer graphene (Fig. 2c [6]), using a measurement geometry similar to Fig. 2a.

Monolayer TMDCs are direct gap semiconductors with spin-valley coupled states in the K and K' valleys, where circularly polarized light excites a particular valley (Fig. 1) [2]. Optical pump-probe measurements established spin-valley lifetimes of a few microseconds in p-type monolayer $WSe_2$ [7,8]. In addition, optical generation of spin-valley polarization in monolayer TMDCs has been used for injecting spins into neighboring graphene layers, which serves as a building block for 2D optospintronics [1]. As shown in Figure 2d [9] circularly-polarized light generates spin-valley polarization in monolayer $MoS_2$, which transfers into graphene, subsequently precesses in a transverse B-field, and is detected by a ferromagnetic electrode. The observation of an anti-symmetric Hanle curve (blue) that flips for opposite detector magnetization (grey) provides convincing evidence for this.

The most recent class of 2D materials for spintronics are the monolayer and few-layer vdW magnets. Intrinsic ferromagnetism was observed in exfoliated $CrI_3$, $CrGeTe_3$, and $Fe_3GeTe_2$ by magneto-optic Kerr effect (MOKE) below room temperature [3]. Room temperature intrinsic ferromagnetism was reported in epitaxial $VSe_2$ and $MnSe_2$, as well as $Fe_3GeTe_2$ modified by patterning or ionic liquid gating [3]. The use of 2D magnets for spintronics (see Section 6) was demonstrated in recent experiments. Electric control of magnetic interlayer coupling was realized in bilayer $CrI_3$, which has a split hysteresis loop indicating antiferromagnetic coupling [3]. As shown in Figure 2e [10], the variation of the splitting with applied electric field indicates real-time control of the antiferromagnetic coupling. Vertical transport through insulating bilayer $CrI_3$ produces a large tunneling magnetoresistance (> 10,000%) due to spin filtering effects [3]. Figure 2f [11] shows the tunneling current as a function of applied magnetic field, showing larger (smaller) current in the parallel (antiparallel) magnetization state. More traditional metal/barrier/metal magnetic tunnel junctions (MTJs) were realized in $Fe_3GeTe_2$/hBN/$Fe_3GeTe_2$ with TMR of 160% [3]. Also relevant for spintronic memory, spin-orbit-torque (see Section 4) was observed in $Fe_3GeTe_2$/Pt [12,13].

*Current and future challenges.* While 2D magnets exhibit a range of interesting magnetoelectronic phenomena, these have only been observed at low temperatures. So far, none of the room temperature 2D ferromagnets have exhibited high remanence or ability to integrate into heterostructures. Thus, continued materials development is needed to simultaneously increase Curie temperature, magnetic remanence, material integration capability, and air-stability. A recent advance along these lines is a Fe-rich version of $Fe_3GeTe_2$, namely $Fe_5GeTe_2$ [14] which exhibits ferromagnetic order near room temperature. Further work on developing new room temperature 2D magnets with improved characteristics is an important challenge.

For applications in magnetoelectronic memories, one of the challenges of the field is to reduce the critical current density needed for magnetic switching. Three viable approaches are spin-orbit-torque in FM/heavy metal bilayers (see Section 4), spin-transfer-torque in FM/barrier/FM MTJs (see Section 6), and voltage controlled magnetism. 2D magnets are attractive in this regard, as the strong covalent bonding of the atomic sheets enables low magnetic volume by scaling down to atomic layers. Reported values of critical current densities for spin-orbit-torque switching in initial studies is ~$10^{11}$ A/m$^2$ [12,13], which is promising. Further development with alternative heavy metal layers such as $WTe_2$, $Bi_2Se_3$ and other vdW materials with high spin-orbit coupling should improve device performance. Strong electrostatic gating effects are a hallmark of 2D materials, which will likely maximize effects such as voltage-controlled magnetic anisotropy (VCMA), which is a candidate for low power dynamic magnetization switching [15]. Combinations of VCMA and spin-torque could enable ultra-efficient magnetization switching. For higher switching speed, antiferromagnetic materials such as $MnPS_3$ and other layered trichalcogenides could provide fast switching due to high magnetic resonance frequencies, which is a general motivation for antiferromagnetic spintronics.

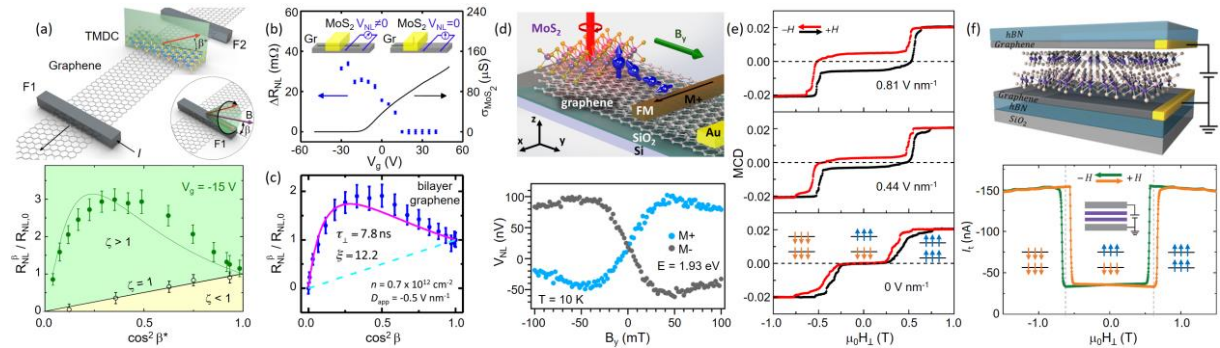

**Figure 2.** (a) Oblique spin precession measurements of TMDC/graphene spin valves, demonstrating proximity spin-orbit coupling through observation of spin lifetime anisotropy. Adapted from [4]. (b) Two-dimensional field-effect spin switch composed of $MoS_2$ on graphene spin valve. Adapted from [5]. (c) Oblique spin precession measurements of dual-gated bilayer graphene spin valves, demonstrating electric control of spin lifetime anisotropy. Adapted from [6]. (d) Opto-valleytronic spin injection from $MoS_2$ into graphene. Adapted from [9]. (e) Electric field control of antiferromagnetic interlayer coupling in bilayer $CrI_3$. Adapted from [10]. (f) Giant spin-filtering tunneling magnetoresistance in vertical transport across bilayer $CrI_3$. Adapted from [11].

For multifunctional spintronics, a crucial issue is understanding and optimizing spin proximity effects in heterostructures of graphene, TMDCs, and 2D magnets. Proximity spin-orbit coupling has been observed in TMDC/graphene and proximity exchange fields have been observed in TMDC/FM insulator and graphene/FM insulator systems [1]. Future challenges include the control of such proximity effects by electric gates and by twist angle between the layers. The ramifications of such proximity effects are in four areas: electrically-controlled spin switches, efficient magnetization switching by spin-orbit torque (see Section 4), optospintronics and optomagnetic switching (see Section 5), and the realization of topological states such as the quantum anomalous Hall effect (QAHE).

*Advances in science and technology to meet challenges.* While exfoliated films are good for fundamental science, epitaxial films are needed for a manufacturable technology. Various forms of chemical vapour deposition have been useful for growth of graphene and TMDCs, while molecular beam epitaxy has been useful for growth of 2D magnets and TMDCs. Optimizing such materials and controlling interface quality is crucial in many contexts. To maximize spin proximity effects, it is important to employ methods for achieving clean interfaces, such as the stacking of 2D materials inside gloveboxes or under vacuum. For many air-sensitive 2D conductors and magnets, stacking inside a glovebox is essential. Electrical spin injection into graphene requires injection across tunnel barriers, which continues to advance.

The use of advanced microscopies and spectroscopies that can image the magnetic order and electronic structure will be important for developing new 2D magnets and spintronic heterostructures. Spin-polarized scanning tunneling microscopy can image magnetism with atomic resolution to correlate the atomic-scale structure with the magnetic ordering, as discussed by D. Sander in the 2017 Magnetism Roadmap. NV diamond microscopy can probe the local magnetic field of buried layers with high spatial resolution. Second harmonic generation is a nonlinear optical probe that is sensitive to symmetry-breaking, which therefore probes the layer stacking and antiferromagnetic order. Micron and nanometer-scale angle-resolved photoemission spectroscopy (micro/nanoARPES) enables spatial mapping of electronic band structure, which will be important for the development of 2D magnets, topological edge states, and spintronic devices.

*Concluding remarks.* The study of spin and magnetism in vdW heterostructures is in its early stages and progressing rapidly, as exemplified by the recent emergence of spin proximity effects and 2D magnets. The development of electrically-tunable, multifunctional spintronic devices will rely on coupled advances in synthesis, assembly, and measurement and will take advantage of the unique properties of 2D materials.